\documentstyle[12pt,twoside,fleqn,espcrc1]{article}

\newcommand{\AmS}{{\protect\the\textfont2
  A\kern-.1667em\lower.5ex\hbox{M}\kern-.125emS}}
\newcommand{\ba}{\begin{eqnarray}}
\newcommand{\ea}{\end{eqnarray}}
\title{Spectrum generating algebra of the symmetric top}

\author{R. Bijker\address{Instituto de Ciencias Nucleares, U.N.A.M., 
        A.P. 70-543, 04510 M\'exico, D.F., M\'exico}
        and
        A. Leviatan\address{Racah Institute of Physics, The Hebrew University, 
        Jerusalem 91904, Israel}}

\begin{document}
% typeset front matter
\maketitle

\begin{abstract}
We consider an algebraic treatment of a three-body system.
We develop the formalism for a system of three identical
objects and show that it provides a simultaneous description of the
vibrational and rotational excitations of an oblate symmetric top.
\end{abstract}

\section{INTRODUCTION}

The study of few-body problems has played an important role in 
many areas of physics \cite{fbs}.
Over the years accurate methods have
been developed to solve the few-body equations. The degree of
sophistication required depends on the physical system, {\it i.e.}
to solve the few-body problem in atomic physics requires a far
higher accuracy than in hadronic physics.

In recent years the development and application of algebraic methods
to the many-body problem ({\it e.g.} collective excitations in nuclei
\cite{ibm} and molecules \cite{vibron}) has received considerable
attention. In spectroscopic studies these algebraic methods
provide a powerful tool to study symmetries and selection
rules, to classify the basis states, and to calculate matrix elements.
In this paper we discuss an application of algebraic methods to the
few-body problem. Especially in the area of hadronic physics, which
is that of strong interactions at low energies, for which exact
solutions of QCD are unavailable, these methods may become very useful 
\cite{BIL}. 

\section{THREE-BODY SYSTEM}

The internal motion of a three-body system can be described in
terms of Jacobi coordinates, $\vec{\rho}$ and $\vec{\lambda}$,
which in the case of three identical objects are
$\vec{\rho} = (\vec{r}_1 - \vec{r}_2)/\sqrt{2}$ and 
$\vec{\lambda} = (\vec{r}_1 + \vec{r}_2 -2\vec{r}_3)/\sqrt{6}$.
Instead of a formulation in terms of coordinates and momenta 
we use second quantization by introducing 
a dipole boson with $L^P=1^-$ for each independent
relative coordinate, and an auxiliary scalar boson with $L^P=0^+$
\ba
p^{\dagger}_{\rho,m} ~, \; p^{\dagger}_{\lambda,m} ~, \;
s^{\dagger} \hspace{1cm} (m=-1,0,1) ~. \label{bb}
\ea
The scalar boson does not represent an independent degree of freedom,
but is added under the restriction that the total number of bosons
$N=n_{\rho}+n_{\lambda}+n_s$ is conserved. This procedure leads to a
spectrum generating algebra of $U(7)$. 
The model space is spanned by the symmetric irreducible 
representation $[N]$ of $U(7)$, which contains the oscillator 
shells with $n=n_{\rho}+n_{\lambda}=0,1,\ldots,N$.
The value of $N$ determines the size of the model space.

\section{PERMUTATION SYMMETRY}

For three identical objects ({\it e.g.} for X$_3$ molecules or
nonstrange $qqq$ baryons) the Hamiltonian (or mass operator) has to be
invariant under the permutation group $S_3$.
The eigenstates of a $S_3$ invariant Hamiltonian are characterized
by the irreducible representations of $S_3$. 
However, in anticipation of the discussion of the oblate top we
prefer to use a labeling under the point group $D_3$ (which is
isomorphic to $S_3$): $A_1$ and $A_2$ for the one-dimensional
symmetric and antisymmetric
representations, and $E$ for the two-dimensional representation.
The scalar boson $s^{\dagger}$ transforms as the
symmetric representation $A_1$, whereas the two dipole bosons
$p^{\dagger}_{\rho,m}$ and $p^{\dagger}_{\lambda,m}$ transform as
the two components of the mixed-symmetry representation,
$E_{\rho}$ and $E_{\lambda}$, respectively.

The Hamiltonian is expressed in terms of the creation and annihilation 
operators of Eq.~(\ref{bb}), such that the total number of bosons 
$N$ is conserved. All one- and two-body $S_3$ invariant interactions 
have, in addition to angular momentum $L$, parity $P$ and permutation 
symmetry $t$, still another symmetry. They commute with 
\ba 
\hat F_2 &=& -i \sum_{m} \left( p^{\dagger}_{\rho,m} p_{\lambda,m}
- p^{\dagger}_{\lambda,m} p_{\rho,m}  \right) ~. 
\ea
This gives rise to an extra quantum number $M_F$, which has a direct 
connection with the permutation symmetry $t$ \cite{BIL,BDL} and 
plays a role similar to that of the label $g$ 
in the context of a six-dimensional harmonic oscillator model \cite{KM} 
and the label $G$ discussed in \cite{Watson} for $X_3$ molecules. 
The corresponding $U(7)$ eigenstates are then labeled by
\ba
\left| [N],\alpha,M_F,L^P_t \right> ~,
\label{bas1}
\ea 
where $\alpha$ denotes the extra labels which are needed for a 
unique classification of the states. 

\begin{figure}[t]
\centering
\setlength{\unitlength}{1.0pt}
\begin{picture}(300,120)(0,0)
\thicklines
\put ( 25, 10) {\circle*{10}}
\put (125, 10) {\circle*{10}}
\put ( 75,110) {\circle*{10}}
\put ( 25, 10) {\line ( 1,0){100}}
\put ( 25, 10) {\line ( 1,2){ 50}}
\put (125, 10) {\line (-1,2){ 50}}

\put (175, 10) {\circle*{10}}
\put (275, 10) {\circle*{10}}
\put (225,110) {\circle*{10}}
\put (175, 10) {\line ( 3, 2){ 50}}
\put (275, 10) {\line (-3, 2){ 50}}
\put (225,110) {\line ( 0,-1){ 67}}
\end{picture}
\caption[]{Geometry of a three-body system.}
\label{geometry}
\end{figure}
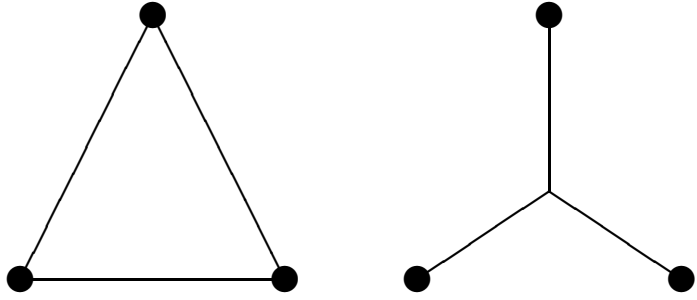

\section{OBLATE SYMMETRIC TOP}

\begin{figure}[t]
\centering
\setlength{\unitlength}{0.8pt}
\begin{picture}(450,175)(0,0)
\thinlines
\put ( 60,  0) {$\nu_1$}
\put ( 25, 50) {\circle*{10}}
\put (125, 50) {\circle*{10}}
\put ( 75,150) {\circle*{10}}
\put ( 25, 50) {\line ( 1,0){100}}
\put ( 25, 50) {\line ( 1,2){ 50}}
\put (125, 50) {\line (-1,2){ 50}}
\multiput (  5, 35)( 5, 0){29}{\circle*{0.1}}
\multiput ( 75,150)( 0,-5){23}{\circle*{0.1}}
\thicklines
\put ( 75,150) {\vector( 0, 1){25}}
\put ( 25, 50) {\vector(-4,-3){20}}
\put (125, 50) {\vector( 4,-3){20}}
\thinlines
\put (210,  0) {$\nu_{2a}$}
\put (175, 50) {\circle*{10}}
\put (275, 50) {\circle*{10}}
\put (225,150) {\circle*{10}}
\put (175, 50) {\line ( 1,0){100}}
\put (175, 50) {\line ( 1,2){ 50}}
\put (275, 50) {\line (-1,2){ 50}}
\multiput (195, 35)( 5, 0){13}{\circle*{0.1}}
\multiput (225,150)( 0,-5){23}{\circle*{0.1}}
\thicklines
\put (225,150) {\vector( 0, 1){25}}
\put (175, 50) {\vector( 4,-3){20}}
\put (275, 50) {\vector(-4,-3){20}}
\thinlines
\put (360,  0) {$\nu_{2b}$}
\put (325, 50) {\circle*{10}}
\put (425, 50) {\circle*{10}}
\put (375,150) {\circle*{10}}
\put (325, 50) {\line ( 1,0){100}}
\put (325, 50) {\line ( 1,2){ 50}}
\put (425, 50) {\line (-1,2){ 50}}
\multiput (315, 30)(   5,2){21}{\circle*{0.1}}
\multiput (365, 50)(1.75,5){21}{\circle*{0.1}}
\thicklines
\put (375,150) {\vector( 1, 0){25}}
\put (325, 50) {\vector(-1,-2){10}}
\put (425, 50) {\vector(-1, 2){10}}
\end{picture}
\caption[]{Schematic representation of the normal vibrations
of a nonlinear X$_3$ configuration. The Jacobi coordinates are indicated
by the dotted lines.}
\label{fund}
\end{figure}
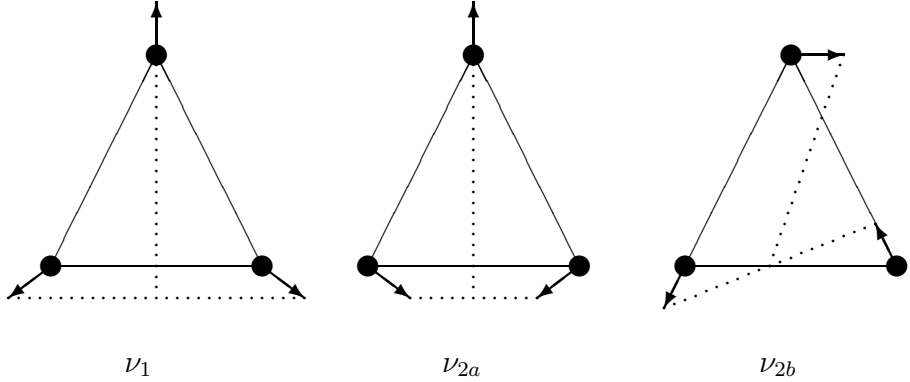

The general one- and two-body $S_3$ invariant Hamiltonian has a 
rich symmetry structure. Here we consider an intrinsic Hamiltonian 
\ba 
H_{\mbox{int}} &=& \xi_1 \, \Bigl ( R^2 \, s^{\dagger} s^{\dagger}
- p^{\dagger}_{\rho} \cdot p^{\dagger}_{\rho}
- p^{\dagger}_{\lambda} \cdot p^{\dagger}_{\lambda} \Bigr ) \,
\Bigl ( h.c. \Bigr )
\nonumber\\
&& +\, \xi_2 \, \Bigl [
\Bigl( p^{\dagger}_{\rho} \cdot p^{\dagger}_{\rho}
- p^{\dagger}_{\lambda} \cdot p^{\dagger}_{\lambda} \Bigr ) \,
\Bigl ( h.c. \Bigr )
+ \Bigl ( p^{\dagger}_{\rho} \cdot p^{\dagger}_{\lambda} 
        + p^{\dagger}_{\lambda} \cdot p^{\dagger}_{\rho} \Bigr ) \,
\Bigl ( h.c. \Bigr ) \Bigr ] ~, \label{hint} 
\ea 
which describes the vibrational excitations of the configuration  
of Fig.~\ref{geometry}, in which the geometry of the three objects 
is that of an equilateral triangle \cite{BIL,BDL}. In the limit 
of a large model space ($N$ large), the vibrational energies are  
\ba 
E_{\mbox{int}} &=& \epsilon_1 \, v_1 + \epsilon_2 \, v_2 ~,
\label{eint}
\ea
with $\epsilon_1 = 4N\xi_1 R^2$ and $\epsilon_2 = 4N\xi_2 R^2/(1+R^2)$. 
Here $v_1$ denotes the number of quanta in the symmetric stretching 
mode ($\nu_1$) with $A_1$ symmetry, and $v_2$ the total number of quanta 
in the antisymmetric stretching ($\nu_{2a}$) and the bending modes 
($\nu_{2b}$), which form a degenerate doublet with $E$
symmetry (see Fig.~\ref{fund}), in agreement with a point group 
classification of the fundamental vibrations for a nonlinear $X_3$ 
configuration \cite{Herzberg}. Hence $H_{\mbox{int}}$ describes the 
vibrations of an oblate symmetric top.

In a geometric description, the excitations of an oblate top
are labeled by 
\ba
\left| (v_1,v_2^l),K,L^P_t \right> ~.
\label{bas2}
\ea
The label $l$ is associated with the degenerate vibration.
It is proportional to the vibrational angular momentum about the axis
of symmetry and can have the values
$l=v_2,v_2-2, \dots,1$ or 0 for $v_2$ odd or even, respectively.
The rotational states, which are characterized by the angular momentum
$L$ and its projection $K$ on the three-fold symmetry axis,
are arranged in bands built on top of each vibration. The projection
$K$ can take the values $K=0,1,2,\dots$~, while the values of the
angular momentum are $L=K,K+1,K+2,\ldots$~.
The parity is $P=(-)^K$, and $t$ denotes the transformation
character of the total wave function under $D_3$.
On the other hand, the eigenstates of the Hamiltonian of Eq.~(\ref{hint}) 
are labeled by Eq.~(\ref{bas1}). A geometric analysis of the eigenstates 
of $H_{\mbox{int}}$ shows, that the label $M_F$ is related to the 
geometric labels $K$ and $l$ of the oblate top \cite{BDL}
\ba 
M_F &=& | K \mp 2l | ~. 
\ea 
For $l=0$ or $K=0$ there is only one value of $M_F$,  
but for $l>0$ and $K>0$ there are two possible values 
of $M_F$ for each $K$. 

\begin{figure}[t]
\centering
\setlength{\unitlength}{0.9pt}
\begin{picture}(400,325)(0,0)
\thinlines
\put (  0, 20) {\line(1,0){400}}
\put (  0,325) {\line(1,0){400}}
\put (  0, 20) {\line(0,1){305}}
\put (400, 20) {\line(0,1){305}}
\thicklines
\put ( 30, 60) {\line(1,0){20}}
\put ( 30, 90) {\line(1,0){20}}
\put ( 30,150) {\line(1,0){20}}
\put ( 30,240) {\line(1,0){20}}
\put ( 80, 96) {\line(1,0){20}}
\put ( 80,156) {\line(1,0){20}}
\put ( 80,246) {\line(1,0){20}}
\multiput (100, 96)(3.75,-2){9}{\circle*{0.1}}
\multiput (100,156)(3.75,-2){9}{\circle*{0.1}}
\multiput (100,246)(3.75,-2){9}{\circle*{0.1}}
\put (130, 80) {\line(1,0){20}}
\put (130,140) {\line(1,0){20}}
\put (130,230) {\line(1,0){20}}
\put (180,158) {\line(1,0){20}}
\put (180,248) {\line(1,0){20}}
\multiput (200,158)(3,-3.2){11}{\circle*{0.1}}
\multiput (200,248)(3,-3.2){11}{\circle*{0.1}}
\put (230,126) {\line(1,0){20}}
\put (230,216) {\line(1,0){20}}
\put (280,246) {\line(1,0){20}}
\multiput (300,246)(2,-3.2){16}{\circle*{0.1}}
\put (330,198) {\line(1,0){20}}
\thinlines
\put ( 40,265) {$\vdots$}
\put ( 30,290) {$M_F=2$}
\put ( 80,290) {$M_F=1$}
\put (130,290) {$M_F=3$}
\put (180,290) {$M_F=0$}
\put (230,290) {$M_F=4$}
\put (280,290) {$M_F=1$}
\put (330,290) {$M_F=5$}
\put ( 35, 35) {$K$=0}
\put (115, 35) {$K$=1}
\put (215, 35) {$K$=2}
\put (315, 35) {$K$=3}
\put ( 52, 60) {$0^+_{E}$}
\put ( 52, 90) {$1^+_{E}$}
\put ( 52,150) {$2^+_{E}$}
\put ( 52,240) {$3^+_{E}$}
\put (102, 96) {$1^-_{E}$}
\put (102,156) {$2^-_{E}$}
\put (102,246) {$3^-_{E}$}
\put (152, 80) {$1^-_{A_1 A_2}$}
\put (152,140) {$2^-_{A_2 A_1}$}
\put (152,230) {$3^-_{A_1 A_2}$}
\put (202,158) {$2^+_{A_1 A_2}$}
\put (202,248) {$3^+_{A_2 A_1}$}
\put (252,126) {$2^+_{E}$}
\put (252,216) {$3^+_{E}$}
\put (302,246) {$3^-_{E}$}
\put (352,198) {$3^-_{E}$}
\end{picture}
\caption[]{Rotational spectrum of a $(v_1,v_2^{l=1})$ 
vibrational band with $t=E$ symmetry.}
\label{vib1}
\end{figure}
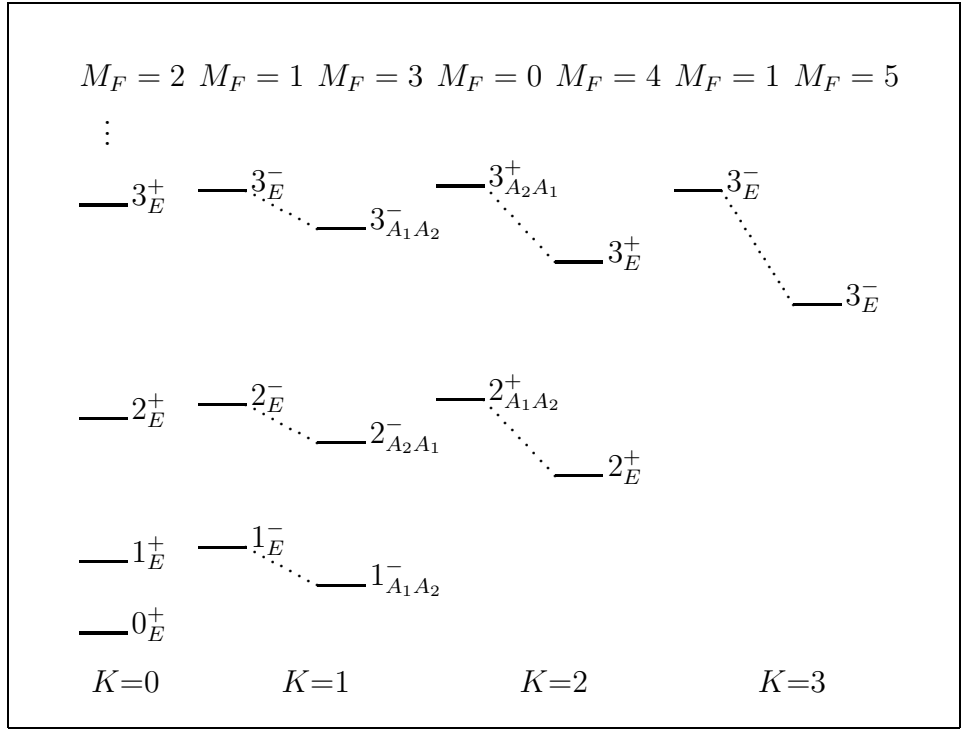

In order to analyze the rotational excitations of an oblate top
we consider a rotational Hamiltonian 
\ba
\hat H_{\mbox{rot}} &=& \kappa_3 \, \hat L \cdot \hat L
+ \kappa_4 \, \hat F_2^2  ~, \label{hrot}
\ea
with eigenvalues 
\ba
E_{\mbox{rot}} &=& \kappa_3 \, L(L+1) + \kappa_4 \, M_F^2
\nonumber\\
&=& \kappa_3 \, L(L+1) + \kappa_4 \, (K^2 \mp 4Kl + 4l^2) ~.
\label{erot}
\ea
The last term contains the effects of the Coriolis force which
gives rise to a $8 \kappa_4 Kl$ splitting between $+l$ and $-l$ 
levels, which increases linearly with $K$.
In Fig.~\ref{vib1} we show the classification scheme for the rotational 
levels with $L \leq 3$ belonging to a $(v_1,v_2^{l=1})$
vibrational band with $E$ symmetry. 
The two $L^{-}_{E}$ levels with $K=3$ are characterized 
by different values of $M_F$~. This shows that $M_F$ provides
an additional quantum number which is needed for a unique 
classification of the states of an oblate top.
According to Eq.~(\ref{erot}) the splitting of the levels with $K=0$
is zero, whereas for $K>0$ there remains a twofold degeneracy because
of the two projections $\pm K$ on the symmetry axis. In particular,
the rotational spectrum of Fig.~\ref{vib1} does not exhibit $l$-type
doubling, which is reflected by the occurrence of degenerate 
doublets of $A_1$ and $A_2$ levels. 
This degeneracy can be lifted by
introducing higher order interactions that break the $M_F$ symmetry.
For example, there exist three-body $D_3$ invariant interactions
that mix states with $\Delta M_F=\pm 6$ \cite{BDL}. 

\section{CONCLUSIONS}

We have presented an algebraic treatment of the three-body problem.
The relative motion of the three-body system is treated by the method
of bosonic quantization, which for the two relative vectors
gives rise to a $U(7)$ spectrum generating algebra. The model
space is spanned by the symmetric irreducible representation
$[N]$ of $U(7)$.

In particular, we have studied the case of three identical objects and
showed how the corresponding permutation symmetry can be taken into
account exactly.
For the special case of one- and two-body interactions, the
eigenstates can be labeled by an additional quantum number $M_F$.
This label plays a very interesting role. On the one hand, it has a
direct connection to the permutation symmetry. On the other hand,
in the limit of a large model space ($N$ large) it is
directly related to the geometric labels $K$ and $l$ of the oblate top,
and provides an extra label which is needed to classify
the rotations and vibrations of the oblate top uniquely.

It was shown that $U(7)$ provides a unified treatment of
both rotational and vibrational excitations of an oblate top.
The ensuing algebraic treatment of the oblate top has found useful
applications both in hadronic physics (nonstrange $qqq$ baryons
\cite{BIL}) and in molecular physics (X$_3$ molecules 
\cite{BDL}). 

Finally, we note that, although we have only discussed the
case of three identical objects, the algebraic procedure is valid in
general, both for three nonidentical objects 
(linear or bent) and for more complicated
systems such as for example the four-body system with tetrahedral 
symmetry. 

\section*{ACKNOWLEDGEMENTS}

This work is supported in part by DGAPA-UNAM under project IN101997,
and by grant No. 94-00059 from the United States-Israel Binational
Science Foundation.

\end{document}